\documentclass[letter,traditabstract]{aa}
\usepackage{graphicx} 
\usepackage{txfonts} 

\begin{document}
\title{On mixing at the core-envelope interface during classical nova
outbursts}

\author{Jordi Casanova \inst{1}
\and Jordi Jos\'e \inst{1}
\and Enrique Garc\'\i a--Berro \inst{2}
\and Alan Calder \inst{3}
\and Steven N. Shore \inst{4} }

\offprints{J. Jos\'e, \email{jordi.jose@upc.edu}}

 \institute{Dept. F\'\i sica i Enginyeria Nuclear, EUETIB,
          Universitat Polit\`ecnica de Catalunya, 
          c/Comte d'Urgell 187, E-08036 Barcelona, Spain,
      \& Institut d'Estudis Espacials de Catalunya, c/Gran Capit\`a 2-4, 
          Ed. Nexus-201, E-08034 Barcelona, Spain\\
	  \and
      Dept. de F\'\i sica Aplicada,
          Universitat Polit\`ecnica de Catalunya, 
          c/Esteve Terrades 5, E-08860 Castelldefels, Spain
      \& Institut d'Estudis Espacials de Catalunya, c/Gran Capit\`a 2-4, 
          Ed. Nexus-201, E-08034 Barcelona, Spain\\
	  \and 
      Department of Physics \& Astronomy, 
          Stony Brook University, 
          Stony Brook, NY 11794-3800\\
	  \and
      Dipartimento di Fisica ``Enrico Fermi'',
          Universit\`a di Pisa and INFN, Sezione di Pisa, 
          Largo B. Pontecorvo 3, I-56127 Pisa, Italy\\
		                    }
 \date{Received February 1, 2010}

\abstract
{Classical novae  are powered by  thermonuclear runaways 
that  occur on
the  white  dwarf component  of  close  binary  systems.  During these 
violent  stellar events,  whose  energy release  is  only exceeded  by
gamma-ray bursts and supernova  explosions, about $10^{-4} - 10^{-5}\,
M_{\sun}$ of material is 
ejected into the interstellar medium. Because of the high
peak temperatures attained during the explosion, $T_{\rm peak} \sim (1
- 4) \times  10^8$ K,  the  ejecta are  enriched in  nuclear-processed
material relative to  solar abundances, containing significant amounts
of $^{13}$C, $^{15}$N, and $^{17}$O  and traces of other isotopes. The
origin  of these  metal enhancements  observed  in the  ejecta is  not
well-known and has puzzled theoreticians for about 40 years.
 In this paper,
we  present  new  2-D  simulations  of  mixing  at  the  core-envelope
interface. We  show that Kelvin-Helmholtz  instabilities can naturally
lead  to self-enrichment  of  the solar-like  accreted envelopes  with
material from the outermost layers of the underlying white dwarf core,
at levels that agree with observations.}

\keywords{(Stars:) novae,  cataclysmic  variables  --  nuclear  reactions,
         nucleosynthesis, abundances --   convection   --
         hydrodynamics -- instabilities -- turbulence }

\titlerunning{Mixing during classical nova outbursts} 
\authorrunning{J. Casanova et al.} 

\maketitle

\section{Introduction}

The assumption of spherical symmetry  in classical nova models (and in
general, in stellar explosions)  excludes an entire sequence of events
associated  with  the way that a  thermonuclear  runaway (hereafter,  TNR)
initiates (presumably  as a point-like ignition)  and propagates.  The
first study of localized TNRs on white dwarfs was carried out by Shara
(1982) on the  basis of semianalytical models. He  suggested that heat
transport was too inefficient to  spread a localized TNR to the entire
white  dwarf surface, concluding  that localized,  {\it volcanic-like}
TNRs were likely to occur.   But his analysis, based only on radiative
and conductive transport, ignored  the major role played by convection
on the lateral thermalization of a TNR.

The importance  of multidimensional effects  for TNRs in  thin stellar
shells was revisited by Fryxell \& Woosley (1982). In the framework of
nova outbursts,  the authors concluded  that the most  likely scenario
involves TNRs  propagated by small-scale  turbulence. On the  basis of
dimensional  analysis  and  flame  theory,  the  authors  derived  the
velocity  of  the deflagration  front  spreading  through the  stellar
surface, in the form $v_{\rm  def} \sim (h_p v_{\rm conv} / \tau_{\rm
burn})^{1/2}$,  where  $h_p$ is  the  pressure  scale height,  $v_{\rm
conv}$ the  characteristic convective velocity,  and $\tau_{\rm burn}$
the characteristic timescale for fuel burning. Typical values for nova
outbursts  yield $v_{\rm  def} \sim  10^4$ cm  s$^{-1}$ (that  is, the
flame propagates halfway throughout the stellar surface in about $\sim
1.3$ days). Shear-driven mixing induced by accretion of matter 
possessing angular momentum was also investigated by Kutter \& Sparks
(1987),
but their numerical simulations failed to obtain a strong enough TNR
to power a nova outburst (see Sparks \& Kutter 1987).

The first  multidimensional hydrodynamic  calculations of this process 
were 
performed by 
 Shankar,  Arnett \& Fryxell  (1992) and Shankar \&  Arnett (1994).
They evolved an accreting,  $1.25 \, M_{\sun}$ white dwarf 
with a 1-D hydro code that was  mapped into a 2-D domain (a
spherical-polar
grid of 25$\times$60 km).  The  explosive event was then followed with
a 2-D  version of  the Eulerian code  {\it PROMETHEUS}.   A 12-isotope
network, ranging from  H to $^{17}$F, was included  to treat the
energetics of  the explosion.   Unfortunately, the subsonic  nature of
the problem, coupled with the use of an explicit code (with a timestep
limited  by  the   Courant-Friedrichs-Levy  condition),  posed  severe
limitations on the  study, which had to be  restricted to very extreme
(rare) cases, characterized by huge temperature perturbations of about
$\sim  100  - 600$\%,  in  small regions  at  the  base of  the
envelope. The total computed time was only about 1
second.    The  calculations   revealed  that   instantaneous,  local
temperature  fluctuations  cause  Rayleigh-Taylor  instabilities.
Their 
rapid rise  and subsequent expansion (in a  dynamical timescale) cools
the hot material  and halts  the lateral  spread of  the burning
front, suggesting that such local temperature fluctuations are not
important 
in the initiation or early stages of the TNR.   The study, therefore,
favored  the local
volcanic-like TNRs proposed by Shara (1982).

Glasner  \&  Livne  (1995) and  Glasner,  Livne \&  Truran
(1997; GLT97) revisited  these early  attempts using  2-D simulations 
performed  with  the  code  {\it VULCAN},  an  arbitrarily  Lagrangian
Eulerian (ALE)  hydrocode capable of handling  both explicit and
implicit steps. As in Shankar et  al. (1992), a slice of the star (0.1
$\pi^{rad}$), in spherical-polar  coordinates with reflecting boundary
conditions, was adopted.  The  resolution near the envelope base was
around 5$\times$5 km.  As before, the evolution of an accreting, $1 \,
M_{\sun}$  CO white dwarf  was initially  followed using  a 1-D
hydro code (to overcome  the early, computationally challenging phases
of  the TNR),  and  then  mapped into  a  2-D domain  as  soon as  the
temperature at  the envelope base  reached $T_{\rm b} \sim  10^8$ K.
As in the previous works, the 2-D runs relied on a 12-isotope network.
The  simulations showed  a  good agreement  with  the gross  picture
described by 1-D models (for instance, the critical role played by the
$\beta^+$-unstable nuclei $^{13}$N,  $^{14,15}$O, and $^{17}$F, in the
ejection  stage, and consequently,  the presence  of large  amounts of
$^{13}$C,  $^{15}$N,  and  $^{17}$O  in the  ejecta).   However,  some
remarkable  differences  were  also  identified.  The  TNR  was
initiated  by  an ensemble of  irregular, localized  eruptions  at  the
envelope base caused  by buoyancy-driven temperature
fluctuations indicating 
that  combustion proceeds in a host of many localized
flames  -- not  as   a  thin   front --  each   surviving  only   a
few
seconds. Nevertheless, these  authors concluded that turbulent
diffusion
efficiently dissipates any local burning around the core, so 
the  fast stages of  the TNR  cannot be  localized and  the
runaway must spread through  the entire envelope. In contrast to 
1-D models, the core-envelope  interface was convectively unstable,
providing a source for  the metallicity enhancement of the envelope
by means of a
Kelvin-Helmholtz   instability  ---  resembling the convective
overshooting  proposed  by  Woosley
(1986).  Efficient  dredge-up of  CO material  from  the outermost
white dwarf layers accounts for  $\sim 30$\% metal enrichment of the
envelope (the accreted envelope  was assumed to be solar-like, without
any   pre-enrichment),  in  agreement   with  the  inferred
metallicites  in  the  ejecta   from  CO  novae (Gehrz et al. 1998). 
Finally, larger
convective eddies were observed,  extending up to  2/3 of the envelope
height with
typical velocities  $v_{\rm conv} \sim 10^7$  cm s$^{-1}$.  Despite
these differences, however, the expansion
and  progress  of  the  TNR  towards the  outer  envelope  quickly
became almost
spherically symmetric, although the initial burning process was not.

The results of  another set of 2-D simulations were published shortly
afterward by
Kercek,  Hillebrandt \&  Truran (1998; KHT98), which  aimed to confirm the
general behaviors  reported by GLT97, in  this case
with a 
version  of the  Eulerian {\it  PROMETHEUS} code.  A similar
domain (a  box of  about 1800$\times$1100 km)  was adopted,  but using
a
cartesian,  plane-parallel  geometry to  allow  the  use of  periodic
boundary  conditions.  Two  resolution simulations were performed,  one
with a
coarser 5$\times$5 km  grid as in GLT97, and a second
with a  finer 1$\times$1 km  grid.  The calculations used
the  same  initial  model  as  GLT97,  and  produced 
qualitatively  similar but  somewhat  less violent  outbursts.
In particular, they obtained  
longer TNRs  with lower  $T_{\rm peak}$  and  $v_{\rm ejec}$,
caused by  large differences in the convective  flow patterns.  Whereas
GLT97 found that a few, large convective eddies
dominated the
flow, most  of the early  TNR was now  governed by small,  very stable
eddies  (with $l_{\max} \sim$  200  km),  which  led to  more  limited
dredge-up   and   mixing  episodes.   The   authors  attributed  these 
discrepancies to  the different  geometry and, more  significantly, to
the boundary conditions adopted in both simulations.

The  only 3-D  nova  simulation to  date  was performed  by
Kercek,  Hillebrandt  \&  Truran  (1999),   adopting  a
computational  domain  of   1800$\times$1800$\times$1000  km  with  a
resolution   of  8$\times$8$\times$8   km.  It  produced flow
 patterns that were
dramatically different  from those found in the  2-D simulations (much
more erratic in the 3-D  case), including mixing by turbulent motions 
occurring
on very small scales (not fully resolved with the adopted resolution)
and peak  temperatures being achieved that were slightly lower  than in  the 2-D
case (a consequence  of the slower and more  limited dredge-up of core
material). The envelope  attained a maximum  velocity that
was  a  factor  $\sim  100$  smaller than  the  escape  velocity  and,
presumably, no mass ejection  (except for a possible wind
mass-loss phase). In view of these results, the authors concluded that
CO mixing must take place prior  to the TNR, in contrast to the main
results  of  GLT97\footnote{Other
multidimensional  studies (Rosner  et al.  2001, 
Alexakis et  al. 2004a,b) focused on  the role of  shear instabilities in
the stratified  fluids that form  nova envelopes. They  concluded that
mixing can  result from  the resonant interaction  between large-scale
shear flows  in the accreted envelope  and gravity waves  
at the interface between the envelope and the underlying white dwarf.
However, to account  for significant mixing, a  very high
shear (with a specific velocity profile) had to be assumed.}.

In summary,  two independent studies,   GLT97
and KHT98, based upon the same  1-D initial model, reached
nearly opposite  conclusions about  the strength of  the runaway  and
its
capability to power  a fast nova. The origin  of these differences was
carefully analyzed  by Glasner, Livne \& Truran  (2005), who concluded
that the early stages of the explosion, prior to the onset of the TNR
-- 
when the evolution is almost quasi-static --  are extremely sensitive
to the outer boundary conditions (see e.g.,  Glasner, Livne \& Truran
(2007), for a  2-D nova simulation  initiated when the temperature  at the
envelope  base is  only $5  \times  10^7$ K).   Several outer  boundary
conditions   were  examined.  The   study  showed   that  Lagrangian
simulations, where the envelope is allowed to expand and mass is
conserved, are consistent  with spherically symmetric
solutions. In contrast, in
Eulerian schemes with a ``free outflow''  outer boundary
condition 
--- the choice adopted in KHT98 ---, the outburst can be artificially quenched.

In light of these conundrums, a  reanalysis of  the role  of late
mixing  at the  core-envelope  interface during  nova outbursts  seems
mandatory. To this end, we  performed an independent  2-D
simulation, identical
to GLT97 and KHT98, with  another
multidimensional   hydrodynamic  code to  investigate
whether mixing can occur in  an Eulerian framework with an appropriate
choice of the  outer boundary conditions. 

\section{Models and input physics}

The 2-D simulation reported in  this paper used {\it FLASH}, a
parallelized, hydrodynamical, Eulerian code
based on the piecewise parabolic interpolation of physical quantities
for solving the  hydrodynamical equations and with an adaptive mesh
refinement procedure. FLASH also uses a monotonicity constraint (rather 
than artificial viscosity) to control oscillations near discontinuities, 
a feature shared with the MUSCL scheme of van Leer (1979). 
For consistency  with GLT97 and
KHT98,  the same initial  model was  used. 
The model was  computed by  GLT97 on the  basis of  a 1-D,
implicit  hydro  code, assuming  accretion  of  solar composition  matter
($Z=0.02$) onto the  surface of a $1 \, M_{\sun}$ CO  white dwarf at a
rate  of $5  \times 10^{-9}\,  M_{\sun}$ yr$^{-1}$.  The accumulation
of
matter in degenerate conditions drives a temperature increase in
the  envelope, resulting in a superadiabatic temperature gradient 
and eventually convective  transport.  The
initial  model corresponds  to the  time when  the temperature  at the
innermost envelope zone is $\approx 10^8$  K. At this stage, the mass
of
the   accreted   envelope   reaches  $2   \times   10^{-5}\,
M_{\sun}$. This  radial profile has  been mapped onto a  2-D cartesian
grid of 800$\times$800 km and  is initially relaxed to guarantee
hydrostatic equilibrium. The  initial computational grid
comprises 112 radial layers (including the outermost part of the CO core)
and 512 lateral layers. 
Calculations rely on the adaptive mesh refinement with a minimum 
resolution 1.6$\times$1.6 km (simulations  with a finer 
resolution will be presented in a forthcoming publication).

A reduced  nuclear reaction network was used to  compute the
energetics  of  the explosion:  it  consists  of  13 isotopes  ($^1$H,
$^{4}$He,   $^{12,13}$C,    $^{13,14,15}$N,   $^{14,15,16,17}$O,   and
$^{17}$F --  as in  GLT97  and KHT98
-- 
supplemented  with  $^{18}$F  to  include  the  important  $^{17}$O(p,
$\gamma$)$^{18}$F  reaction),  linked  through  a net  of  18  nuclear
processes (mainly,  p-captures and $\beta^+$-decays).   Reaction rates 
are taken  from Angulo et al. (1999) and some more recent updates (see 
Jos\'e, Hernanz, \& Iliadis 2006, Jos\'e \& Shore 2008, and references therein).

Periodic  boundary  conditions  were   adopted  at  both lateral sides,
while hydrostatic   boundary  conditions   are  fixed   at both    the
bottom
(reflecting) and  the top  (outflow).\footnote{Technical details of
how 
boundary conditions are implemented in Godunov-type codes can be found
in Zingale et  al. (2002).} The set of  boundary conditions are similar
to  those  implemented  in  GLT97 and KHT98, but note that the outer computational
grid  adopted  in GLT97 is Lagrangian  instead  of
Eulerian (to  follow the late  expansion stages of the  TNR). Finally,
energy  transport is  included using an  effective thermal
diffusion   coefficient  that   includes   radiative  and   conductive
opacities (Timmes 2000).

\section{Results}

In GLT97, significant numerical noise
was  present  at  the   onset  of  their  calculations  that produced  
temperature  fluctuations of  about  10--20\%.  We introduced (just at the
initial time-step) a
Gaussian temperature perturbation  at the  core-envelope interface of  5\%. 
For
comparison, the value in KHT98 was 1\%. The size of the
initial perturbation was 
2 km,  much smaller than the  contemporary depth of  the accreted
envelope ($\sim800$
km).  The initial perturbation produces fluctuations that move along
the
core-envelope   interface   during    the   first   seconds   of   the
simulations.  These fluctuations, in turn, spawn Kelvin-Helmholtz
vortices, which clearly show up
about 200 s later (Fig. 1), and appear to initiate a turbulent cascade.
Filaments and buoyant plumes are fully resolved in these simulations.
At this stage, the fluid  is characterized by 
a large Reynolds  number, 
with a characteristic eddy length of 50 km, fluid velocities  of $10^5 -
10^6$ cm s$^{-1}$, and a dynamic viscosity of $10^4$ P.
These  Kelvin-Helmholtz (KH) instabilities 
transport unburnt  CO-rich material from  the outmost
layers of  the white  dwarf core and  inject it into  the envelope.
The characteristic eddy turnover time is l$_{\rm conv}$/v$_{\rm
conv}$ $\sim$ 10 s. 

\begin{figure*}
\centering
\includegraphics[width=0.38\textwidth]{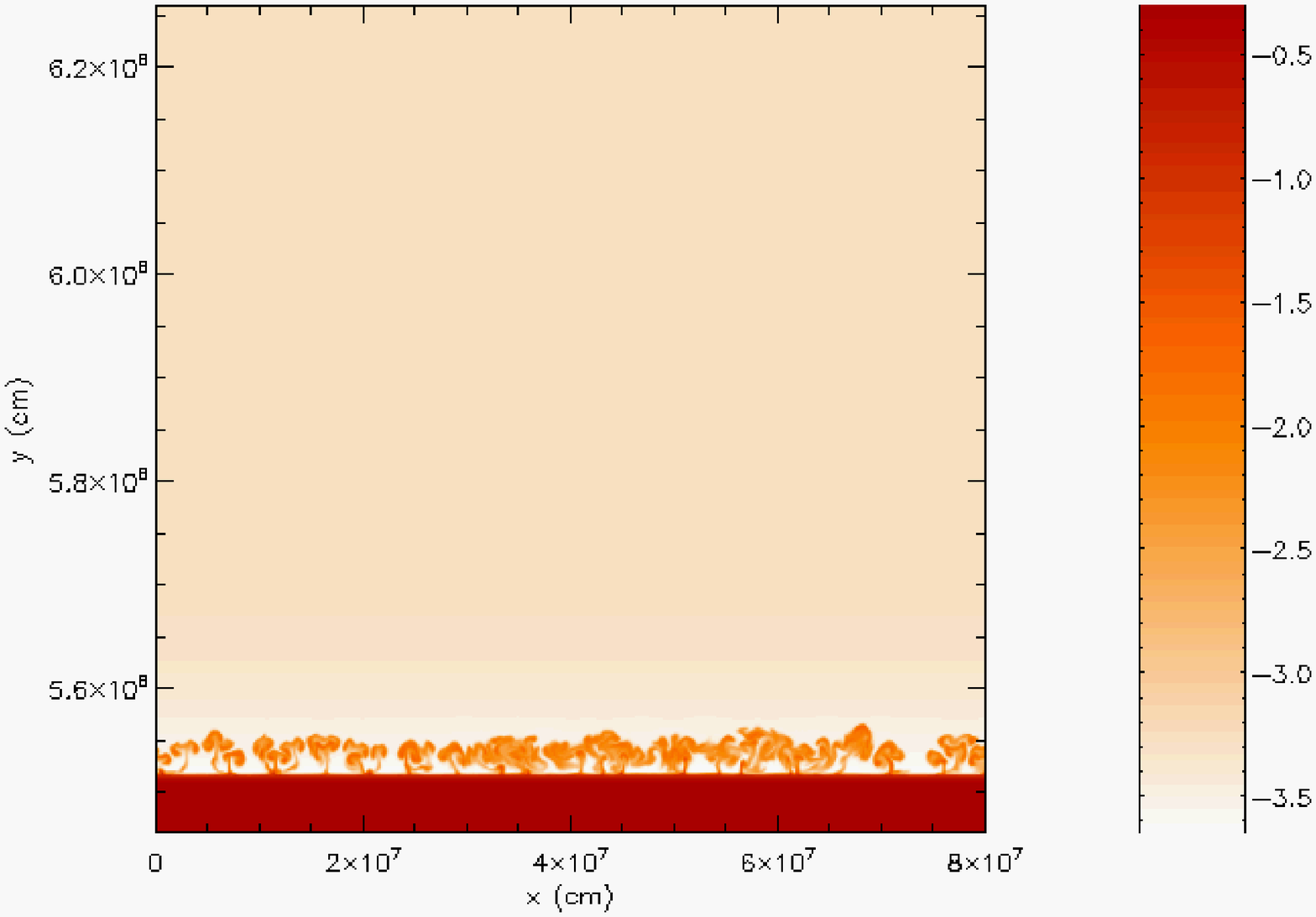}
\includegraphics[width=0.38\textwidth]{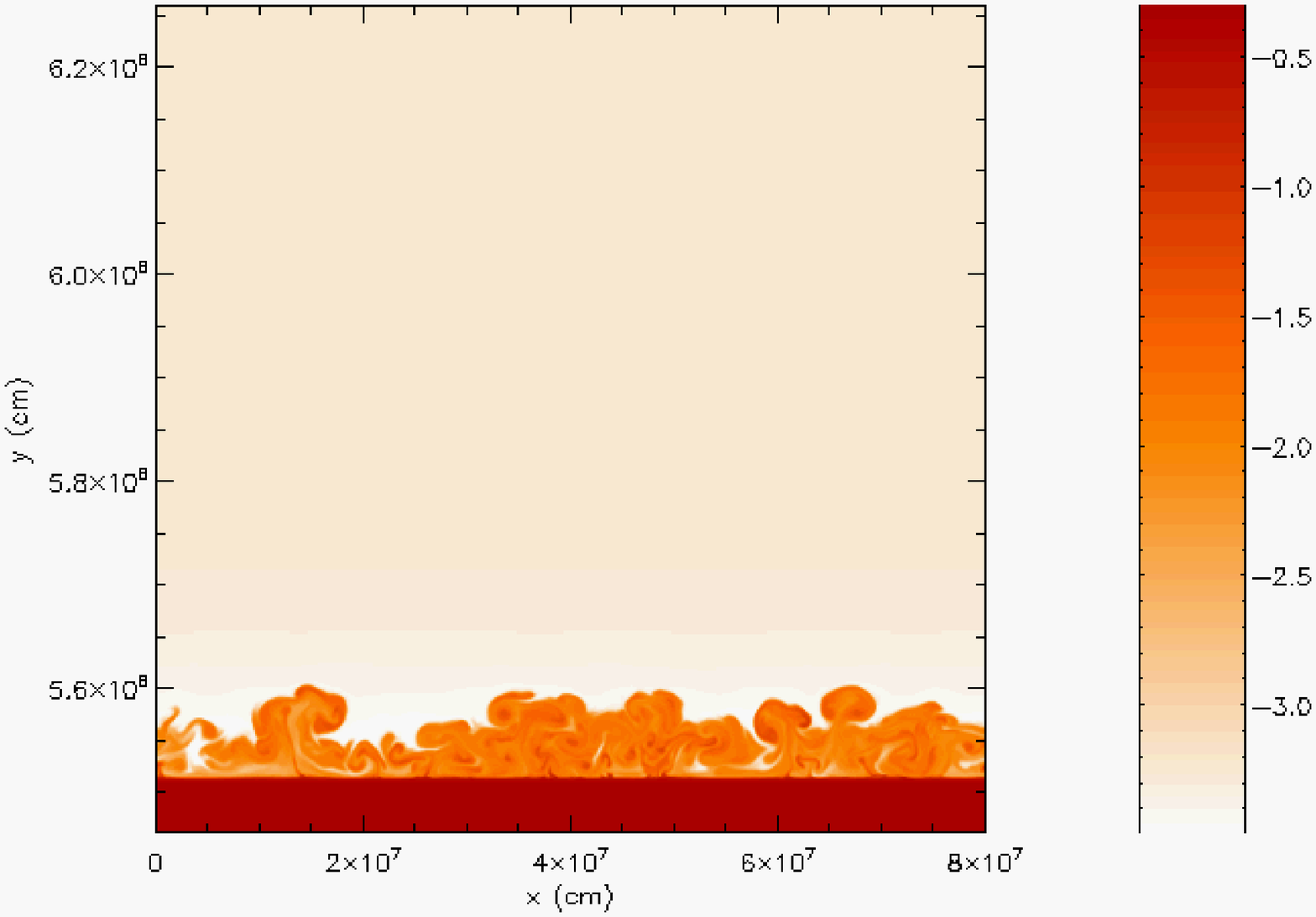}
\includegraphics[width=0.38\textwidth]{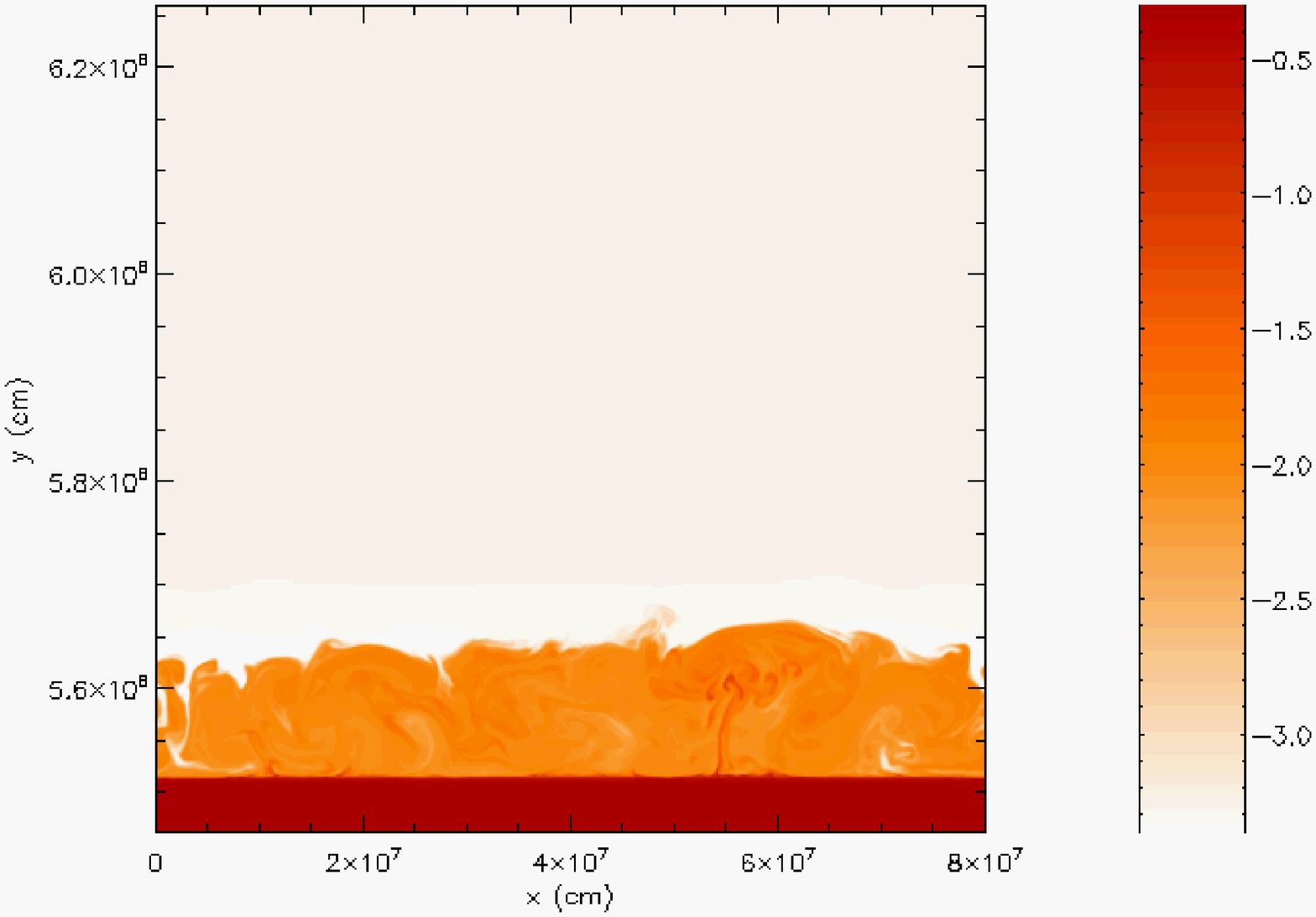}
\includegraphics[width=0.38\textwidth]{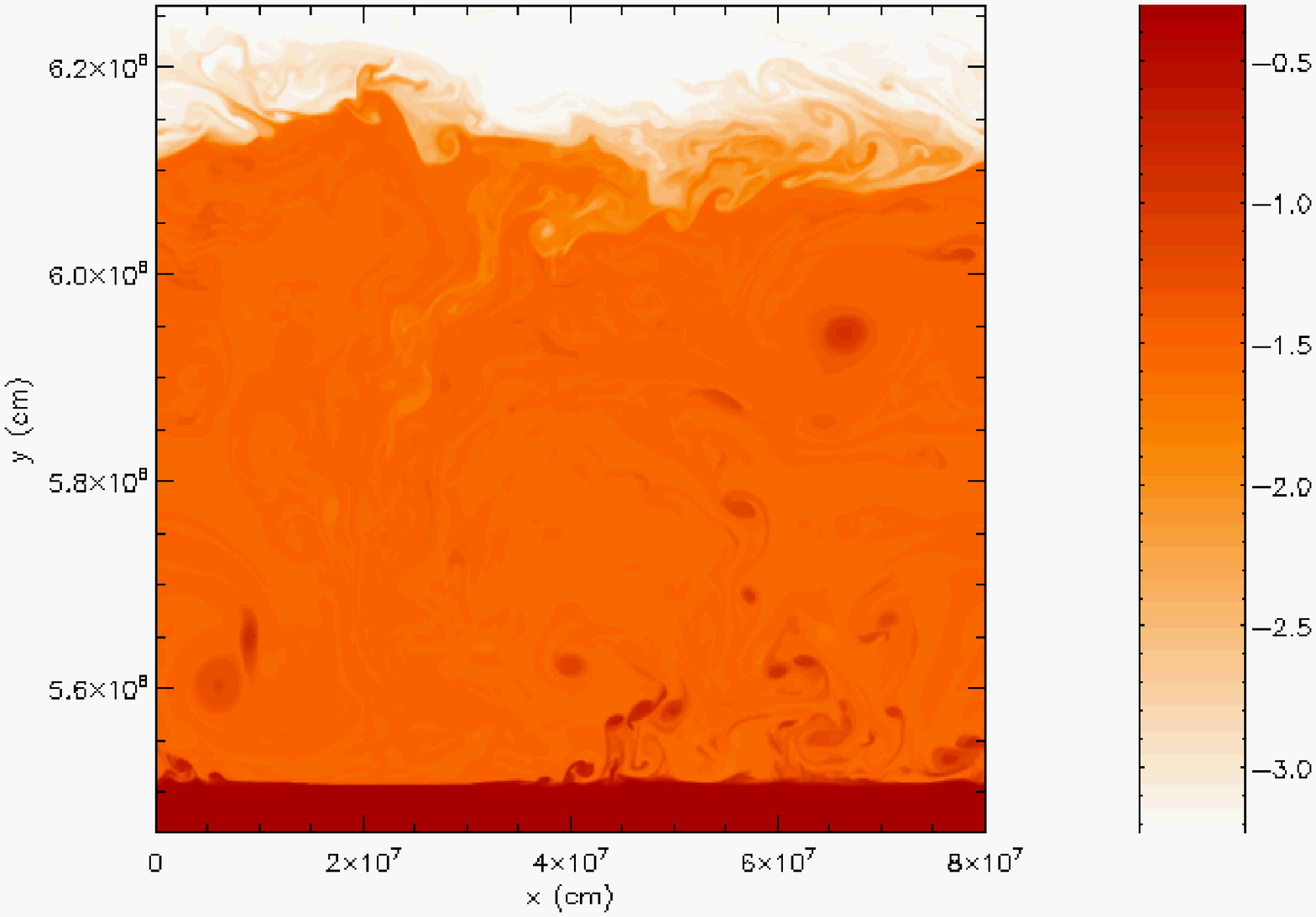}
\caption{Snapshots    of   the   development    of   KH
         instabilities at $t=215$ s  (upper left panel), 235 s (upper
         right), 279 s  (lower left), and 498 s  (lower right), shown
         in terms  of $^{12}$C mass  fraction (in logarithmic scale). 
	  The injection  of core
         material   driven  by  the   KH  instabilities
         translates into  a mass-averaged abundance  of CNO-nuclei in
         the   envelope   of   0.079,   0.082,   0.089,   and   0.17,
         respectively.  The mean  CNO  abundance at  the  end of  the
         simulations reaches 0.30, by mass.}
\label{fig:C12}
\end{figure*}

As  the  KH vortices  grow  in  size, more  CO-rich
material  is transferred   into  the   envelope.    Convection  becomes
 more
turbulent. The initially  small convective eddies  merge into
huge shells (Fig. 2), as seen also in GLT97.  At this  stage,  the nuclear  energy generation  rate
reaches  $10^{15}$  erg g$^{-1}$  s$^{-1}$,  while the  characteristic
burning  timescale decreases  to $\sim$5  s.  The  convective
filaments 
continue  growing in  size  and progressively  occupy  the whole
envelope
length. 
 Although  not resolved  in these  simulations, and in contrast to 
the 3-D case, the conservation of vorticity in 2-D forces
the largest eddies to grow in an inverse vorticity cascade, while energy
flows to the viscous scale
with a distribution that deviates from the Kolmogorov spectrum 
(see e.g., Lesieur, Yaglom, \& David 2000; Shore 2007).
At this time, the  temperature at the envelope base reaches $\sim2
\times 10^8$  K, at fluid velocities of  $10^8$ cm
s$^{-1}$ (of the  order of the escape velocity,  characteristic of the
dynamic phases of the explosion), and a nuclear energy generation rate
of $10^{16}$  erg g$^{-1}$ s$^{-1}$.  The  convective turnover time is
now  $\sim$  5 s.  The  mean CNO
abundance in the envelope has increased to 0.30, a value that agrees
well with 
both the previous  simulations by GLT97 and the
mean  metallicities inferred  from  observations of  the
ejecta in non-neon (CO) novae (see Jos\'e \& Shore 2008).
At this stage, since the outer envelope
layers   had started to   escape  the   computational   (Eulerian)  domain,
simulations were stopped.

\begin{figure*}
 \centering
 \includegraphics[width=0.34\textwidth]{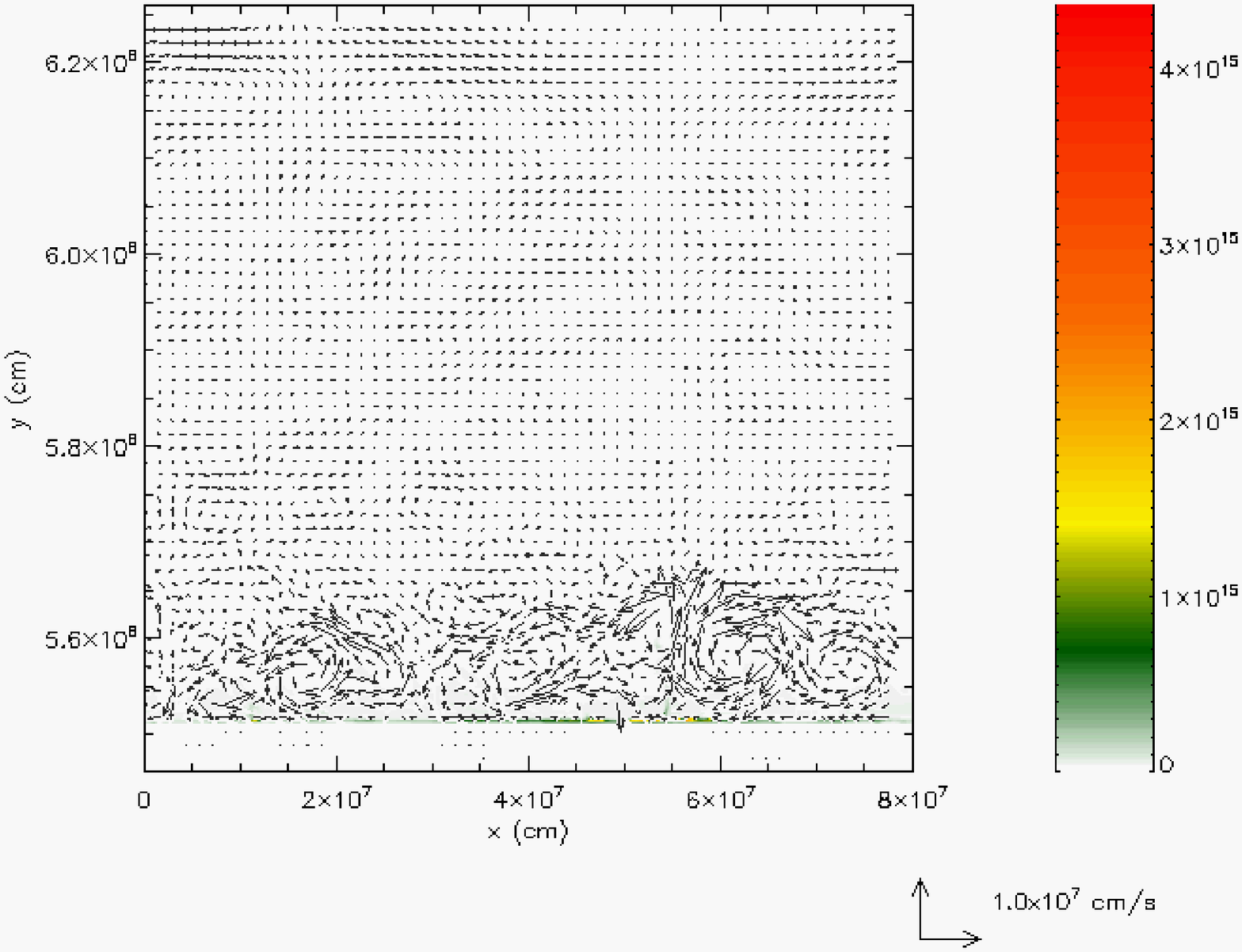}
 \includegraphics[width=0.34\textwidth]{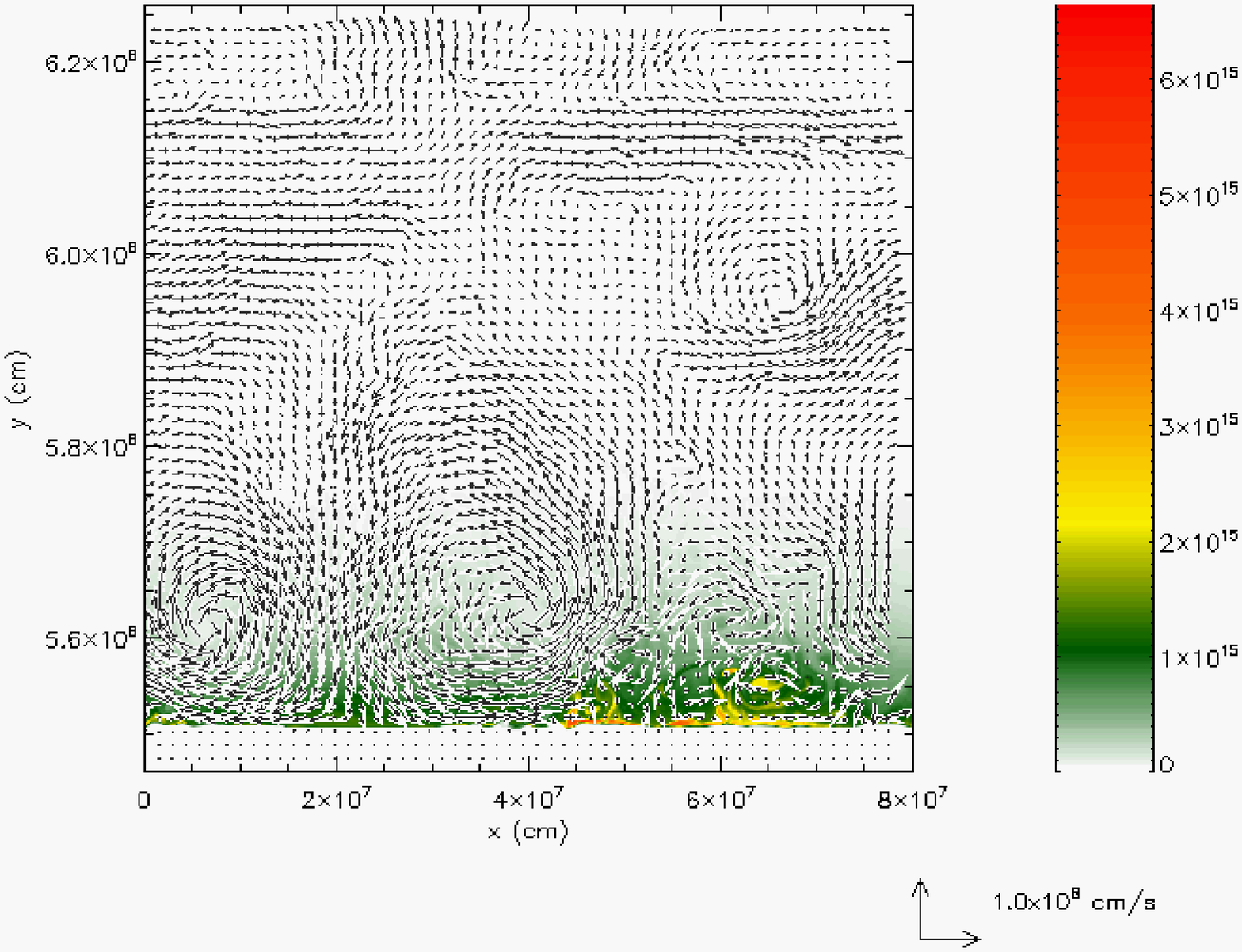}
 \includegraphics[width=0.29\textwidth]{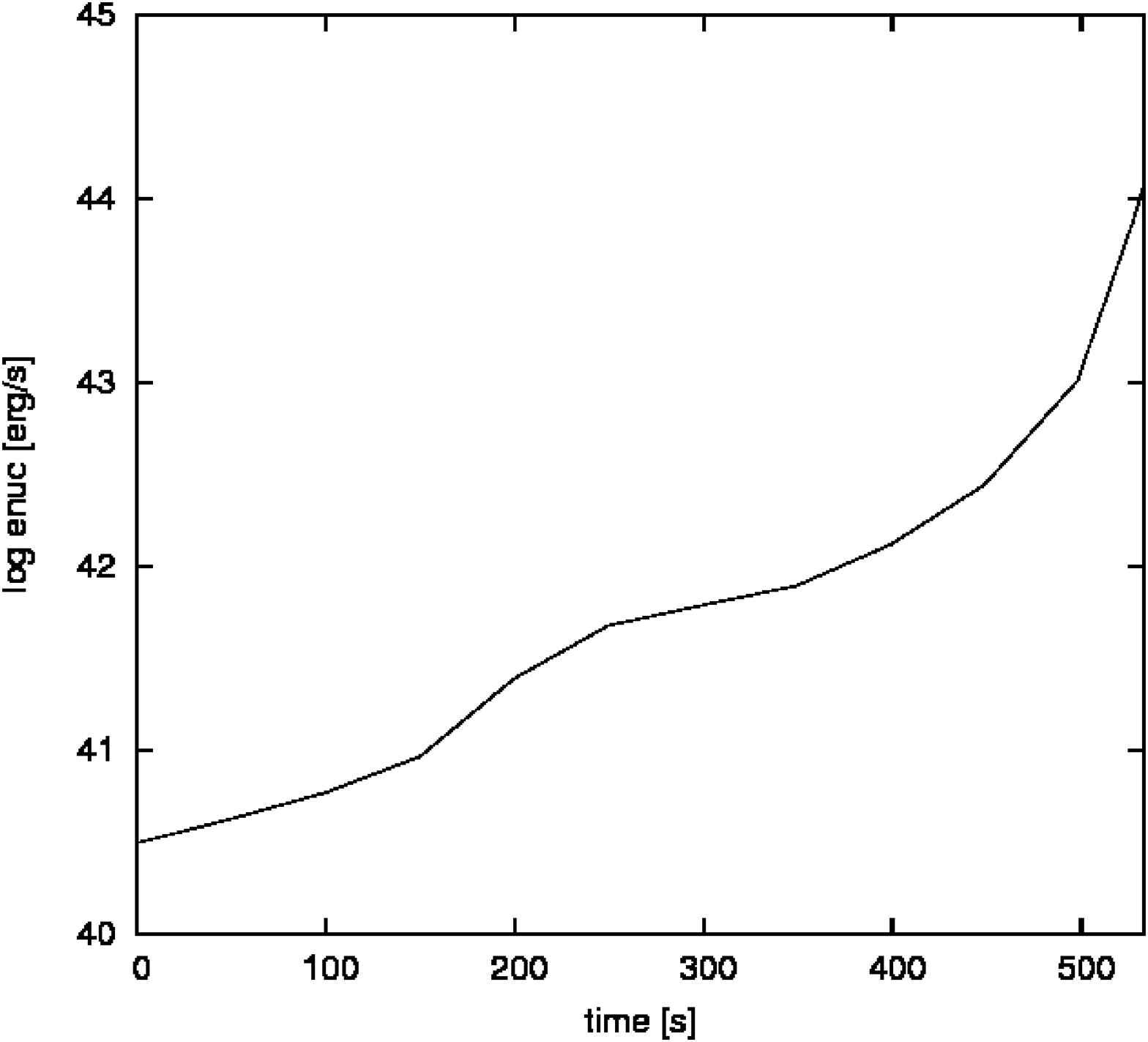}
 \caption{Left and central panels: 
 same as Fig. 1, but  for the velocity fields at t = 279 s (left) 
          and 498 s (right), superimposed on a plot of the nuclear 
	  energy generation rate (in ergs g$^{-1}$ cm$^{-1}$).
	  Right panel: Time evolution of the overall nuclear burning rate.}
\label{fig:Vfield}
\end{figure*}

Our 2-D  simulations, in  agreement with the results  reported in
GLT97, show that the progress and extension of the TNR
throughout  the  envelope occurs with almost  spherical  symmetry,
even though the structure of the ignition  is  not.  This explains
the success  of 1-D  models in  reproducing the gross observational properties
(light curves, velocities of the ejecta, nucleosynthesis) of
nova explosions
(Starrfield  et al.  1998, 2009;  Kovetz  \& Prialnik  1997; Yaron  et
al. 2005; Jos\'e \& Hernanz 1998).

\section{Discussion and Conclusions}

We have analyzed the possible self-enrichment  of the
solar-composition accreted  envelope with material from  the underlying
white
dwarf during nova outbursts in a multidimensional framework.  We have found
that a shear flow at the  core-envelope interface (which unlike the 
spherically symmetric case, does not behave like a rigid wall) 
drives mixing through KH instabilities.  Large
convective eddies develop close to the core-envelope interface, of a
size comparable to the height  of the envelope (similar to 
the pressure scale height in 1-D simulations),
 mixing CO-rich
material  from  the outermost  layers  of  the  underlying white dwarf
into  the  accreted envelope.  The  metallicity enrichment 
achieved in  the envelope,  $Z \sim 0.30$,  is in agreement  with
observations  of CO  nova  ejecta.  Our 2-D simulations also show that
even for a point-like TNR ignition,  the expansion and progress of the
runaway is  almost spherically symmetric  for nova conditions.  We note
that  the  adopted  resolution as  well  as  the  size, intensity, and
location of the initial perturbation have a very limited
impact  on the  results, principally affecting the  timescale 
for the onset of  the KH instability but not the
final, mean  metallicity. Details will be extensively  discussed in a
forthcoming publication.   Our results agree  with earlier
2-D  hydrodynamic simulations  (GLT97)  and  solve the
controversy raised  by another  2-D study (KHT98) that
questioned  the efficiency  of this  mixing mechanism,  and  hence the
corresponding strength  of the runaway  and its capability to  power a
fast nova outburst.

\begin{acknowledgements}  
We greatly appreciate comments and suggestions
from D. Garc\'\i a--Senz, J. Isern, A. Parikh, and M. Zingale on an 
early version of this manuscript.
This work has been partially supported by the Spanish
MEC  grants AYA2007-66256  and AYA2008-04211-C02-01,  and by  the E.U.
FEDER funds.  We also  acknowledge the Barcelona Supercomputing Center
for a generous allocation of time at the Mare Nostrum supercomputer.
\end{acknowledgements}

\end{document}